\documentclass[amsmath,reprint,amssymb,apsan]{revtex4-1}
\renewcommand{\v}[1]{\ensuremath{\mathbf{#1}}} 
\newcommand{\gv}[1]{\ensuremath{\mbox{\boldmath$ #1 $}}} 
\renewcommand{\d}[2]{\frac{d #1}{d #2}} 
\newcommand{\pd}[2]{\frac{\partial #1}{\partial #2}} 
 
\newcommand{\grad}[1]{\gv{\nabla} #1} 
\let\baraccent=\= 
\renewcommand{\=}[1]{\stackrel{#1}{=}} 

\newcommand{\p}{^{\prime} }

\usepackage{graphicx}
\usepackage{dcolumn}
\usepackage{multirow}
\usepackage{natbib}
\usepackage{xcolor}
\usepackage{mathtools}
\usepackage[normalem]{ulem}
\begin{document}
\def\reff@jnl#1{{\rm#1}} 

\def\araa{\reff@jnl{ARA\&A}}             
\def\aj{\reff@jnl{AJ}}                   
\def\apj{\reff@jnl{ApJ}}                 
\def\apjl{\reff@jnl{ApJ}}                
\def\apjs{\reff@jnl{ApJS}}               
\def\apss{\reff@jnl{Ap\&SS}}             
\def\aap{\reff@jnl{A\&A}}                
\def\aapr{\reff@jnl{A\&A~Rev.}}          
\def\aaps{\reff@jnl{A\&AS}}              
\def\baas{\ref@jnl{BAAS}}               
\def\jcap{\reff@jnl{J. Cosmology Astropart. Phys.}}
\def\jrasc{\reff@jnl{JRASC}}             
\def\memras{\reff@jnl{MmRAS}}            
\def\mnras{\reff@jnl{MNRAS}}             
\def\na{\reff@jnl{New A}}                
\def\nar{\reff@jnl{New A Rev.}}          
\def\pra{\reff@jnl{Phys.~Rev.~A}}        
\def\prb{\reff@jnl{Phys.~Rev.~B}}        
\def\prc{\reff@jnl{Phys.~Rev.~C}}        
\def\prd{\reff@jnl{Phys.~Rev.~D}}        
\def\pre{\reff@jnl{Phys.~Rev.~E}}        
\def\prl{\reff@jnl{Phys.~Rev.~Lett.}}    
\def\pasa{\reff@jnl{PASA}}               
\def\pasp{\reff@jnl{PASP}}               
\def\pasj{\reff@jnl{PASJ}}               
\def\qjras{\reff@jnl{QJRAS}}             
\def\physrep{\reff@jnl{Phys.~Rep.}}   
\let\astap=\aap
\let\apjlett=\apjl
\let\apjsupp=\apjs
\let\applopt=\ao
\preprint{APS/123-QED}
\title{Exploring A Cosmic-Ray Origin of the Multi-wavelength Emission in M31 }
\author{Alex McDaniel}
 \email{alexmcdaniel@ucsc.edu}
\author{Tesla Jeltema}
 \email{tesla@ucsc.edu}
 \author{Stefano Profumo}
 \email{profumo@ucsc.edu}
\affiliation{
Department of Physics, University of California,
1156 High Street, Santa Cruz, California, 95064, USA
}
\affiliation{
Santa Cruz Institute for Particle Physics,
1156 High Street, Santa Cruz, California, 95064, USA
}
\date{\today}
\begin{abstract}
A recent detection of spatially extended gamma-ray emission in the central region of the Andromeda galaxy (M31) has led to several possible explanations being put forth, including dark matter annihilation and millisecond pulsars. Another possibility is that the emission in M31 can be accounted for with a purely astrophysical cosmic-ray (CR) scenario. This scenario would lead to a rich multi-wavelength emission that can, in turn, be used to test it. Relativistic cosmic-ray electrons (CRe) in magnetic fields produce radio emission through synchrotron radiation, while X-rays and gamma rays are produced through inverse Compton scattering. Additionally, collisions of primary cosmic-ray protons (CRp) in the interstellar medium produce  charged and neutral pions that then decay into secondary CRe (detectable through radiative processes) and gamma-rays. Here, we explore the viability of a CR origin for multi-wavelength emission in M31, taking into consideration three scenarios: a CR scenario dominated by primary CRe, one dominated by CRp and the resulting secondary CRe and  gamma rays from neutral pion decay, and a final case in which both of these components exist simultaneously. We find that the multi-component model is the most promising, and is able to fit the multi-wavelength spectrum for a variety of astrophysical parameters consistent with previous studies of M31 and of cosmic-ray physics. However, the CR power injection implied by our models exceeds the estimated CR power injection from typical astrophysical cosmic-ray sources such as supernovae.
\end{abstract}

\pacs{Valid PACS appear here}
\maketitle

\section{\label{sec:intro}Introduction}
The study of gamma rays in galactic environments offers an intriguing probe of many physical phenomena including cosmic-ray production and transport, star formation rates, or new physics such as dark matter. The Andromeda galaxy (M31) is particularly enticing as a target of gamma-ray studies due its status as the nearest large spiral galaxy. M31 has been the focus of several previous gamma-ray searches \cite{Fichtel, Pollock, Sreekumar, Blom, Abdo2010, FermiM31, M31Halo, feng}. Early observations  \cite{Fichtel, Pollock, Sreekumar, Blom} were only able to place upper limits until the galaxy was first detected in gamma-rays using 2 years of Fermi-LAT data at $5.3\sigma$ significance, along with some evidence of a spatial extension at the $1.8\sigma$ confidence level \cite{Abdo2010}. It has also been observed by high energy Cherenkov telescopes, though no detection has yet been made at energies above the TeV  \cite{HAWC_M31, Bird2015, Aharonian, Rubenzhal}.

Recently, M31 was detected in gamma-rays by the Fermi telescope at a significance of nearly $10\sigma$ with a detection of spatially extended emission out to $\sim 5$ kpc at the $4\sigma$ significance level \cite{FermiM31}. This emission resembles to some extent the well studied Galactic Center Excess (GCE) of gamma rays in the center of the Milky Way, and has led to comparisons in possible origins for the emission in the two galaxies. Proposed explanations for the GCE include signals of annihilating dark matter \cite{abazajian2014, calore, daylan, GordonMacias,HooperGoodenough, goodenough}, an unresolved population of millisecond pulsars (MSP) \cite{bartels, brandt,EcknerMSP}, or additional cosmic-ray sources \cite{carlson, gaggero15, cholis}. Due to some similarities between both the two galaxies themselves and the observed emissions, it is natural to also consider whether these are viable explanations for the M31 detection. Although there is significant uncertainty in the dark matter density profile, the possibility of a dark matter signal in M31 has previously been studied for gamma-rays \cite{Fornengo, HAWC_M31, li}, as well as other wavelengths \cite{McDaniel2018, M31USC, BeckColaM31, ChanM31, Ng, Boyarsky, Watson}. A brief argument is presented in the recent Fermi detection paper \cite{FermiM31} using the relative $J$-factors of the Galactic center and M31 to infer that the expected gamma-ray emission from dark matter annihilation in M31 is roughly a factor of $\sim 5$ below the observed emission. In a recent paper \cite{McDaniel2018}, we studied the possibility of a dark matter origin of the M31 emission from a multi-wavelength perspective. We found that when assuming a dark matter only interpretation using favored GCE dark matter models, such models typically require annihilation cross-sections above current constraints, and have spectral shapes that are inconsistent with the M31 observations. However, this does not necessarily preclude these models contributing a subdominant component of the observed emission, with the shape of the spectrum predominantly determined by another emission source. Additionally,the dark matter particle models in that analysis that can reproduce the M31 gamma-ray emission also produce synchrotron emission that is in tension with observational radio data. There have also been efforts made to explore a millisecond pulsar (MSP) explanation for the M31 gamma-ray emission \cite{EcknerMSP, FragioneMSP}. Ref. \cite{FragioneMSP} studied MSPs originating from globular cluster disruption in the bulge of M31, whereas Ref. \cite{EcknerMSP} considered MSPs formed {\it in situ}. The {\it in situ} model was found to fit the energetics and morphology of the excess well, however neither study could account for the full detected emission, with each providing only $\sim 1/4$ of the M31 observation.

In light of the lack of a definitive dark matter or exclusive unresolved MSP explanation, this work focuses on the scenario in which this emission can potentially be accounted for using a conventional astrophysical interpretation of cosmic ray induced emissions. In order to explore this possibility, we study the multi-wavelength emission in Andromeda to consider whether a cosmic ray explanation is consistent across the spectrum. We predominantly focus on the radio and gamma-ray emission, as these regimes provide the most insight given currently available observational data.
The production of cosmic rays in astrophysical systems can lead to emissions at various wavelengths from radio, to X-ray and gamma-ray \cite{Blumenthal, longair,rybicki}. When cosmic ray electrons (CRe) are injected into regions containing magnetic fields, they radiate synchrotron emission at radio wavelengths, whereas through inverse Compton scattering the electrons upscatter ambient photons, such as from the CMB or starlight, to X-ray and gamma-ray energies. Gamma rays can also be produced directly through the decay of neutral pions produced in cosmic-ray proton (CRp) collisions with the interstellar medium \cite{dermer, MannheimSchlick1994}. However, as first pointed out in Pshirkov et. al. 2016 \cite{Pshirkov} and also found in the Ackermann et. al. (2017) \cite{FermiM31} analysis, the gamma-ray emission in M31 does not seem to be spatially correlated with neutral gas or regions of high star formation, which are typically the regions wherein CRp are expected to be produced and interact with the ISM to produce the $\pi^0$ gamma-rays. The magnitudes of these fluxes depend on the components of the astrophysical environment, such as magnetic field, target photon radiation field, and abundance of cosmic ray production mechanisms. The diffusion of the relativistic cosmic rays additionally plays a significant role in the expected signal, as cosmic rays diffuse and escape the system, thus suppressing the expected flux. The mechanism by which cosmic rays are produced and accelerated has been a topic of extensive interest. For galaxies, supernovae remnants (SNR) are considered to be the main sources of cosmic rays \cite{ginzburg, berezhko, bell, thompson, drury}. While SNRs are thought to provide the dominant contribution of cosmic rays in galaxies, other mechanisms such as pulsars and their nebula can also provide significant contributions to the total cosmic ray population \cite{AharonianCR, gendelev,hooperPulsar}. The expected power injection from these cosmic ray sources provides a benchmark to which we can compare the cosmic ray power necessary to produce the multi-wavelength emissions in M31.

This paper is organized as follows: in section \ref{sec:astro} we detail the relevant physical modeling, including the magnetic field, interstellar radiation field, and diffusion model, as well as the solution to the diffusion equation. In section \ref{sec:emissions} we present the expressions for each radiative process under consideration, then in section \ref{sec:data} we present the data used in the analysis. In section \ref{sec:results} we present the results of our analysis, in section \ref{sec:xraydiffuse} we comment on X-ray diffuse emission constraints, and, finally, we conclude in section \ref{sec:conclusion}.

\section{\label{sec:astro}Astrophysical Model of Andromeda}
In order to calculate the secondary emission from the production of primary or secondary cosmic-ray electrons, we must first model the relevant astrophysical components of Andromeda. In particular, we require a description of the magnetic field model, which determines the synchrotron emission produced, as well as the inter-stellar radiation field (ISRF) that provides the target photon bath for the inverse Compton (IC) scattering. Also relevant to this analysis is a model of diffusion and radiative energy losses, since the relativistic electrons diffuse out and escape the system, while also losing energy through synchrotron emission, IC emission, Coulomb interactions, and bremsstrahlung. In the following section, we briefly describe the model adopted in this work. 
\subsection{\label{sec:bfield}Magnetic Field}
The study of radio synchrotron emission in M31 requires some knowledge of the magnetic field in our region of interest, namely within the inner few kpc. In this region, the field structure is turbulent and complex, however estimates of the field strengths as determined by Faraday rotation measures of polarized radio emission are typically around $15  \pm 3 \mu$G for $r=0.2-0.4$ kpc, and $19  \pm 3 \mu$G for $r=0.8-1.0$ kpc \cite{hoernes, giessubel}. At larger radii in the disk of the galaxy, the magnetic field falls off to values of roughly $5 \pm 1\mu$G \cite{fletcher}.
In our previous study of Andromeda where we considered a dark matter origin of the multi-wavelength emission we selected a spatially dependent magnetic field based on these values, approximating some degree of spherical symmetry in the central region, while noting that that model would not be applicable at larger radii where a multi-dimensional field model would be more appropriate. In this analysis however, we treat the magnetic field strength as one of the free parameters in our fit, and so we adopt a simplified constant magnetic field where 
\begin{equation}
B(r) = B_{\mu}.
\end{equation}
While this is helpful in that it reduces the parameters we need to fit in order to define our model, it comes at the cost of accuracy in capturing the complexity of the field or any spatial dependence. We can consider the constant field as an average over space, with a consequence of this being that we would expect the values of $B_{\mu}$ to be lower than the quoted central values, and at a roughly similar level to that of the disk.
\subsection{\label{sec:ISRF}Inter-stellar Radiation Field}
Our inter-stellar radiation field model (ISRF) contains two components: (i) a CMB photon component and (ii) a starlight (SL) component. We have chosen to neglect the infrared (IR) component in the ISRF and instead focus only on the CMB and starlight components in order to simplify the ISRF modelling and fitting procedure, implicitly making the assumption that the starlight component will be the more significant contribution to the IC emission. For the CMB, we simply have a black-body spectrum at $T=2.73$ K and spatial homogeneity. Thus we have, 
\begin{equation}\label{n_CMB}
n_{CMB} (\nu) = \frac{8\pi \nu^2}{c^3}\frac{1}{e^{h\nu/kT}-1}.
\end{equation}
For the starlight component, we approximate the spectrum as a black-body with temperature $T=3500$ K, a choice motivated by previous analysis of the ISRF in the Milky Way demonstrating this as a good approximation for starlight spectra. We additionally include a spatial dependence based on the starlight luminosity profile of M31 \cite{courteau} incorporating a bulge component of the form:
\begin{equation}\label{eq:n_bulge}
n_{bulge}(r) \propto e^{-b_n\left[\left(\frac{r}{r_b}\right)^{1/n}-1\right]}.
\end{equation}
and a disk component:
\begin{equation}\label{eq:n_disk}
n_{disk}(r) \propto e^{-\frac{r}{r_d}}.
\end{equation}
Combining these spatial components with the black-body spectral profile yields a starlight photon number density: 
\begin{equation}\label{eq:nSL}
n_{SL}(\nu, r) = N_{SL}\frac{8\pi \nu^2/c^3}{e^{h\nu/kT}-1}\left[ e^{-b_n\left[\left(\frac{r}{r_b}\right)^{1/n}-1\right]}+\frac{e^{-\frac{r}{r_d}}}{135} \right].
\end{equation}
The parameters  $r_b,\: r_d,\:n, \:b_n $ are taken from \cite{courteau} and the factor of $1/135$ in the disk component was chosen to recreate the bulge to disk luminosity ratio in \cite{courteau}. The factor $N_{SL}$ is a dimensionless normalization constant that is to be determined in the later sections as a free parameter in our fit. To get a sense for what value this parameter should be, we can consider the stellar luminosity of the inner region of M31. The stellar luminosity within a 1 kpc radius of M31 has previously been reported as $L = 10^{9.9} L_{\odot}$  \cite{groves, draine}. We can roughly estimate the luminosity as
\begin{equation}
L = 4\pi r^2 c\: \bar{u}_{SL}, 
\end{equation}
where the bar refers to a spatial average over the volume. Taking the radius to be $\sim 1$ kpc, a stellar luminosity of $L = 10^{9.9} L_{\odot}$  corresponds to $\bar{u}_{SL}\approx 5$ eV cm$^{-3}$, or $N_{SL} \approx 5\times 10^{-12}$. A reasonable, albeit somewhat large, range of values for the stellar energy density in the inner regions of galaxies is $\bar{u}_{SL} \approx 1-10$ eV cm$^{-3}$ \cite{ProfumoUllio, porter}, which roughly corresponds to a normalization fit range of $N_{SL} \in (10^{-12}, 10^{-11})$.

\subsection{\label{sec:diff}Solution to the Diffusion Equation}
After being injected into the system, the CRe undergo both radiative losses and diffusion. Diffusion is particularly important on shorter distance scales, such as the few kpc scales considered in this work, and we have demonstrated in our previous M31 paper (see Fig. 6 from \cite{McDaniel2018}) that it significantly impacts the expected fluxes. The diffusion and radiative energy loss mechanisms of the CRe are accounted for in the diffusion equation:
\begin{equation}\label{eq:diffusion}
\begin{split}
\pd{}{t} \d{n_{e^{\pm}}}{E}= &\grad \left [D(E,\v{r}) \grad{ \d{n_{e^{\pm}}}{E} } \right] \\
&+\pd{}{E}\left[b(E,\v{r}) \d{n_{e^{\pm}}}{E} \right]+Q(E,\v{r}).
\end{split}
\end{equation}
where we neglect convection and reacceleration effects which can be safely ignored for energies greater than a few GeV \cite{Delahaye2009, Delahaye2010}. Particularly in a quiescent galaxy such as M31 with its low star-formation rate \cite{Ford2013, Rahmani2016}, the effects of convection are expected to be less prominent than in galaxies with higher star-formation activity \cite{McCormick,Lacki,Veilleux} such as starbursts or even the Milky Way, and thus we treat diffusion as the dominant escape term. In equation \ref{eq:diffusion}, $\partial n_e/\partial E$ is the electron/positron equilibrium spectrum with units of GeV$^{-1}$ cm$^{-3}$, $D(E, \v{r})$ is the diffusion coefficient, $b(E,\v{r})$ is the energy loss term, and $Q(E, \v{r})$ is the CRe source term that we specify in later sections and has units of GeV$^{-1}$ s$^{-1}$ cm$^{-3}$. In the energy loss term, we include contributions from synchrotron, IC, Coulomb, and bremsstrahlung processes, with the full expression given by:
\begin{equation}\label{eq:bloss}
\begin{split}
b(E,\v{r}) & = b_{IC}(E,\v{r}) + b_{Synch.}(E,\v{r}) + b_{Coul.}(E) + b_{Brem.}(E)\\
	& = b_{IC}^0u_{CMB}E^2 + b_{IC}^0u_{SL}(r)E^2 + b^0_{Synch.}B^2(r) E^2 \\
	&+ b_{Coul.}^0 \bar{n}_e\left(1+\log\left(\frac{E/m_e}{\bar{n}_e }\right)/75 \right)\\
	& + b^0_{Brem.}\bar{n}_e \left( \log\left(\frac{E/m_e}{\bar{n}_e }\right) + 0.36 \right).
\end{split}
\end{equation}
The $b^0$ coefficients in this expression have units GeV s$^{-1}$ with values $b_{syn}^0 \simeq 0.0254$, $b_{IC}^0 \simeq 0.76$, $b_{brem}^0 \simeq 1.51$, and $b_{Coul}^0 \simeq 6.13$ \cite{longair, cola}. The photon energy density for the CMB is $u_{CMB} =0.25$ eV cm$^{-3}$ and for the starlight photons can be computed from equation (\ref{eq:nSL}) to be $u_{SL}(r) = h\nu_{0}^2 n_{SL}(\nu_{0}, r)$, where $\nu_0$ is taken to be the peak frequency. Finally, $\bar{n}_e$ in equation (\ref{eq:bloss}) refers to the average thermal electron density and is taken to be  $\bar{n}_e \approx 0.01$ cm$^{-3}$ \cite{Gaensler,BerkMitraMueller,BerkMullerElectron, CaprioliElectron,CaprioliAmatoBlasiElectron}.

For the diffusion coefficient we assume a homogeneous power law of the form:
\begin{equation}\label{eq:Dcoeff}
D(E) = D_0 E^{\delta}
\end{equation}
with $\delta  = 1/3$ and $D_0=3\times10^{28}$ cm$^{2}$ s$^{-1}$ \cite{SMP, vladimirov, BaltzEdsjo, WLG}. The choices of these parameters are motivated by assuming that M31 has roughly similar diffusion properties to the Milky Way, with these values being determined by measurements of the stable (e.g. B/C) or unstable (e.g. Be$^{10}$/Be$^{9}$) secondary to primary ratios, and also supported by studies of the far-infrared - radio correlation in M31 and other galaxies that infer similar values \cite{BerkBeckTab, murphy}. Equation (\ref{eq:diffusion}) can be solved analytically using the Green's function method (see e.g. \cite{cola, ginzburg}) and in the steady state case where the left-hand side of equation (\ref{eq:diffusion}) is set to zero the appropriate Green's function with free-escape boundary conditions is given by:
\begin{equation}
\begin{split}
G(r,  \Delta v ) &= \frac{1}{\sqrt{4\pi \Delta v }}\sum_{n = -\infty}^{\infty} \left(-1\right)^n\int_0^{r_h}dr\p\frac{r\p}{r_n} \left(\frac{ Q(E, r\p) }{Q(E,r)}\right)\\
& \times\left[\exp\left(-\frac{ (r\p - r_n)^2 }{4 \Delta v}	\right)	- 	\exp\left(-\frac{ (r\p + r_n)^2 }{4\Delta v}	\right)	\right],
\end{split}
\end{equation}
where $r_h\approx 5$ kpc is the diffusion zone radius and the locations of the image charges used to implement the free-escape boundary condition are $r_n = \left(-1\right)^nr + 2nr_h$. The value $\Delta v$ is defined as $\Delta v = v(E) - v(E\p)$ with 
\begin{equation}\label{eq:ve}
v(E) = \int_E^{\infty} d\tilde{E} \frac{D(\tilde{E})}{b(\tilde{E})}.
\end{equation}
where we have approximated a spatially independent form of the energy loss term by taking a spatial average of $u_{SL}(r)$ and $B(r)$ in equation (\ref{eq:bloss}). In the above expression, $E\p$ represents the energy of the electron at the source, while $E$ is the interaction energy. The quantity $\sqrt{\Delta v}$ has units of distance, and represents the diffusion length scale of the particles. The final form of the electron equilibrium spectrum is then given by, 
\begin{equation}\label{eq:dndeeq}
\d{n_{e^{\pm}}}{E}\left(E,r\right)= \frac{1}{b(E,r)}\int_E^{\infty}dE\p G\left(r,\Delta v\right)Q(E, r).
\end{equation}
Here we use the full spatially dependent form of the energy loss expression, rather than the homogeneous form used in equation (\ref{eq:ve}).
\section{\label{sec:emissions}Multi-wavelength Emission}
Once we have obtained the electron equilibrium spectrum $dn_{e^{\pm}}/dE$ by solving the diffusion equation, we can then proceed to calculate the emissivity $j_i$, by integrating the electron spectrum with the power for the given radiative process, namely the synchrotron radiation and IC scattering for our purposes. This gives
\begin{equation}\label{eq:emiss}
j_i(\nu, r) = 2\int_{m_e}^{\infty} dE \: P_i(\nu,E,r) \d{n_{e^{\pm}}}{E}\left(E,r\right)
\end{equation}
where the factor of two accounts for electrons and positrons and $P_i$ is the power of a radiative process $i$ which we calculate in the following sections. From here, the flux density is given by the integral of the emissivity over volume,
\begin{equation}
S_{i} (\nu)= \frac{1}{4\pi d^2}\int dV j_i(\nu,r)  \approx \frac{1}{d^2}\int dr \:r^2 j_i(\nu,r) 
\end{equation}
where $d$ is the distance to M31, taken to be $d=780$ kpc \cite{Stanek1998}. In this work, we make use of the publicly available RX-DMFIT tool \cite{RXDMFIT} to solve the differential diffusion equation and then to perform the various secondary emission calculations. Models used in this analysis can be obtained from the authors.
\subsection{\label{sec:synch}Synchrotron Power}
In the presence of ambient magnetic fields, the relativistic CRe undergo synchrotron radiation, producing radio emission. The synchrotron power for a frequency $\nu$ averaged over all directions is \cite{storm16, longair} 
\begin{equation}
P_{syn} \left(\nu, E , r\right) = \int_0^{\pi} d\theta \frac{\sin \theta}{2} 2\pi \sqrt{3}r_0 m_e c \nu_0 \sin\theta F\left(\frac{x}{\sin\theta}\right), 
\end{equation}
where $r_0 = e^2/(m_ec^2)$ is the classical electron radius, $\theta$ is the pitch angle, and $\nu_0 = eB/(2\pi m_e c)$ is the non-relativistic gyrofrequency. The $x$ and $F$ terms are defined as,
\begin{equation}
x \equiv \frac{2\nu m_e^2}{3\nu_0 E^2},
\end{equation}
\begin{equation}
F(s) \equiv s\int_s^{\infty} d \zeta K_{5/3}\left( \zeta \right)\simeq 1.25 s^{1/3}e^{-s}\left[648 + s^2\right]^{1/12},
\end{equation}
where $K_{5/3}$ is the Bessel function of order 5/3. 


\subsection{\label{sec:IC}Inverse Compton Power}
With the photon number density $n(\epsilon, r) = n_{CMB}(\epsilon) + n_{SL}(\epsilon, r)$, and the IC scattering cross-section $\sigma\left( E_{\gamma} , \epsilon , E \right)$, the IC power is

\begin{equation}
P_{IC} \left( E_{\gamma}, E , r\right) = c E_{\gamma}\int d\epsilon \: n \left( \epsilon, r \right) \sigma\left( E_{\gamma} , \epsilon , E \right)
\end{equation}

\noindent
where $\epsilon$ is the energy of the target photons, $E$ is the energy of the relativistic electrons and positrons, and $E_{\gamma}$ is the energy of the photons after scattering (note that $E_\gamma= h\nu$ for observing frequency $\nu$ in equation (\ref{eq:emiss})). The scattering cross-section, $\sigma\left( E_{\gamma} , \epsilon , E \right)$, is given by the Klein-Nishina formula:
\begin{equation}
\sigma\left( E_{\gamma} , \epsilon , E \right) = \frac{3 \sigma_T}{4\epsilon \gamma^2} G\left( q, \Gamma \right),
\end{equation}

\noindent where $\sigma_T$ is the Thomson cross-section and $G(q, \Gamma)$ is given by \cite{Blumenthal}:
\begin{equation}
G (q, \Gamma) = \left[  2q\ln q + (1+2q)(1-q) + \frac{(2q)^2(1-q)}{2(1+\Gamma q)}  \right],
\end{equation}
where,
\begin{equation}
\Gamma = \frac{4\epsilon \gamma}{m_e c^2} = \frac{4\gamma^2 \epsilon}{E}, \qquad q = \frac{E_{\gamma}}{\Gamma(E-E_{\gamma})} 
\end{equation}
The kinematics of inverse Compton scattering set the range of $q$ to be $1/\left( 4 \gamma^2 \right) \leq q \leq1$ \cite{cola, Blumenthal, rybicki}.

\subsection{\label{sec:gamma}Gamma-ray Flux}
In addition to gamma rays produced from IC scattering, we also consider gamma rays resulting from the decay of neutral pions produced in cosmic-ray proton collisions. When the pions are produced they decay rapidly within a time span of $\sim 10^{-16}$ s. The gamma rays do not experience diffusion or radiative loss effects, and thus we do not need to consider equation (\ref{eq:diffusion}). Instead, for a $\pi^0$ gamma-ray source injection $Q_{\gamma}$ (in units of GeV$^{-1}$ cm$^{-3}$ s$^{-1}$), the flux is simply given by integrating over the volume of the source \cite{cola, RegisUllio, McDaniel2018}:
\begin{equation}
F_{\gamma}= \frac{1}{d^2}\int dr  r^2 E^2Q_{\gamma}\left(E, r\right).
\end{equation}

\section{\label{sec:data}Gamma-ray and Radio Data}
The gamma-ray data points are taken from the analysis performed in Ackermann et. al. (2017) \cite{FermiM31}, where they used 88 months of PASS 8 Fermi data collected between August 4, 2008, and December 1, 2015. Reconstructed events within an energy range of 0.1-100 GeV were considered as well as reconstructed directions within a $14^{\circ}\times 14^{\circ}$ region centered at $(\alpha,\delta) = (10^{\circ}.6847, 41^{\circ}.2687)$. SOURCE class events were used excluding those with zenith angle greater than $90^{\circ}$ or rocking angle greater than $52^{\circ}$. The resulting detected emission found in this study was concentrated within the inner 5 kpc, motivating this as the choice of region of interest in our calculations of the gamma-ray emission resulting from IC scattering  and pion decay.

Radio observations of M31 have predominantly focused on regions of large radii out to about $\sim 16$ kpc with a particular emphasis on the star-forming 10 kpc ``ring'' \cite{RadioData, M31_350, BeckBerkWiel, BeckBerkHoernes, BeckGrave}, or alternatively on the central regions within $r\sim 1$ kpc \cite{WalterbosGrave, BerkWielBeck}. For our purposes, the available data in the 1 kpc region are most useful, as they allows us to focus our analysis on the inner region and make better comparisons between the 1 kpc ROI for radio emission and the 5 kpc ROI for gamma-ray emission. In each of the spectral energy distributions in the following sections, the synchrotron emission is calculated with $r=1$ kpc while the IC and gamma-ray emission are calculated with $r = 5$ kpc.

\section{\label{sec:results}Results}
In our analysis, the multi-wavelength emission in M31 is assumed to be due to the presence of cosmic-rays. We consider here two cosmic-rays production mechanisms: the first is primary production of CRe following a power law with exponential cut-off source injection, which then radiate synchrotron and IC emission. The second source of cosmic-rays we consider is primary production of cosmic-ray protons obeying a power law. The hadronic inelastic interactions of the CRp produce neutral pions that decay into gamma rays, as well as charged pions that decay into secondary CRe which then produce synchrotron and IC emission. Finally, we consider the scenario in which both of these sources provide comparable contributions to the overall cosmic-ray abundance in what we refer to as our “multi-component” model. We then examine to what extent each of these three scenarios can be responsible for the multi-wavelength emission in M31. Of the three cases mentioned, the multi-component model appears the most convincing, while the primary-only and secondary-only models do not easily reproduce the emission in M31 within the range of realistic parameter space.
\subsection{\label{sec:resultsCRPrimary}Emission from primary cosmic ray electrons}
We now define the source term of equation (\ref{eq:diffusion}) by considering the case in which the cosmic-ray population is dominated by primary electron production obeying a power law with an exponential cutoff:
\begin{equation}\label{eq:PrimSource}
Q_{e^{\pm}}(E) = N_{CR_e} \left(\frac{E}{\text{GeV} }\right)^{-\alpha_{e}} e^{-E/E_{cut}}.
\end{equation}
In this section, $\alpha_{CR}, \: N_{CR_e}, \: E_{cut}$ in equation (\ref{eq:PrimSource}) along with $B_{\mu}$ and $N_{SL}$ are taken to be free parameters that we adjust to fit the observed radio and gamma-ray spectra. Previous studies of cosmic-ray origins can provide some guidance as to reasonable values for these parameters. For example, values of $\alpha_e \sim 2.0- 2.3$ have been found to be consistent with production of cosmic rays in supernovae (SNe), as well as suggesting $E_{cut}$ values on the order of a few TeV \cite{Delahaye2010, DiMauro2014, manconi2017, Fang2017,bernardo2013}. The normalization $N_{CR_e}$ however is poorly constrained, and in section \ref{sec:CRePower} we compare the fit values of $N_{CR_e}$ with the corresponding SNe power output. 

\begin{figure}[h!]
\centering
\includegraphics[width=0.45\textwidth]{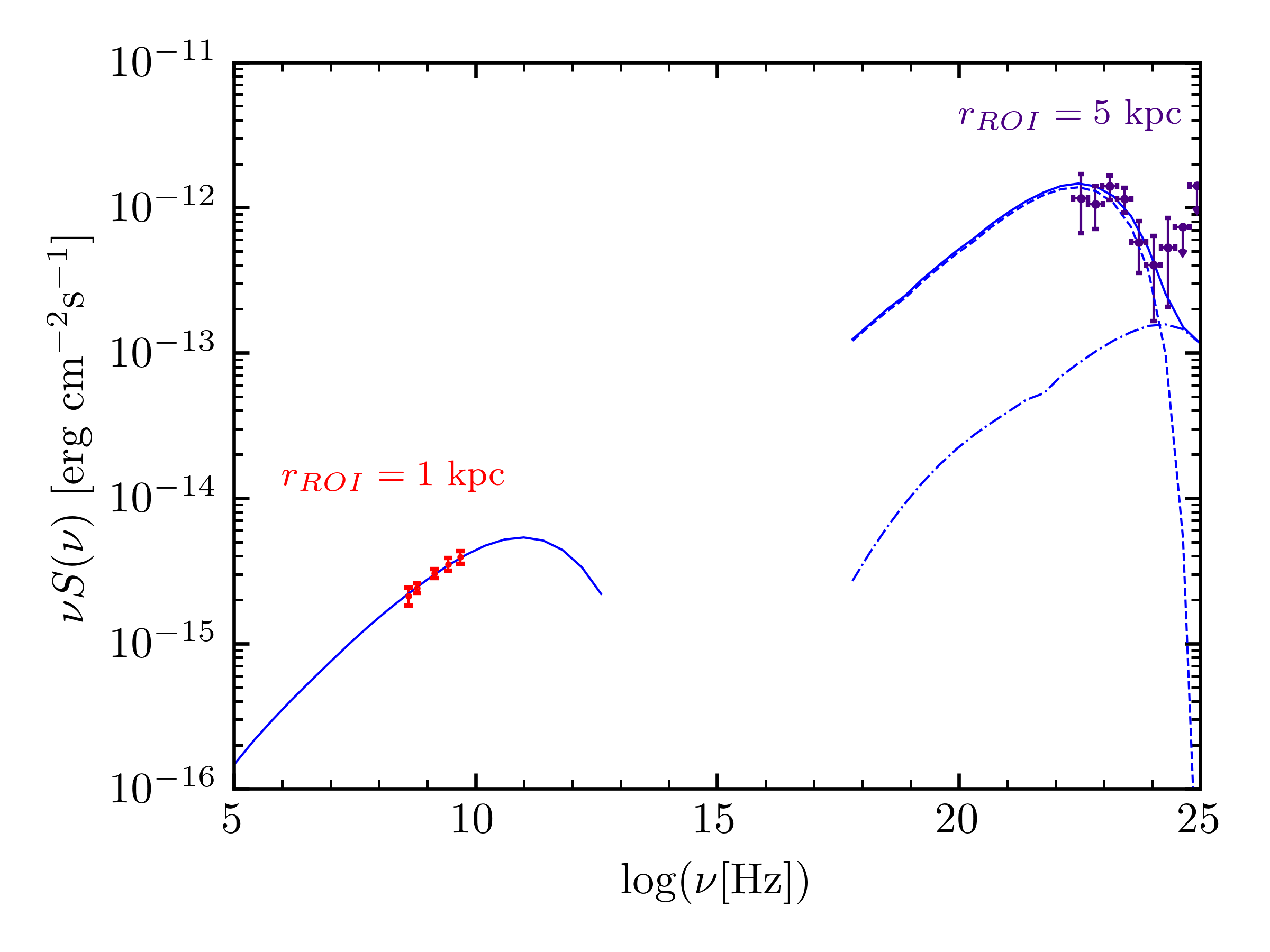}
\hfill
\caption{Synchrotron and IC emission from primary production of CRe for the best fit model in table \ref{tab:PrimFitParams}.  The dashed lines are the CMB IC contribution, the dash-dot are the SL IC contribution, and the total emissions are the solid lines. Radio data are taken from \cite{WalterbosGrave} and gamma-ray data are taken from \cite{FermiM31} }
\label{fig:PrimBF}
\end{figure}

In table \ref{tab:PrimFitParams} we list the results for the best-fitting model and plot the SED in figure \ref{fig:PrimBF}. Throughout this analysis we fit the parameters by minimizing the standard $\chi^2$ metric using a Nelder-Mead simplex algorithm  \cite{NelderMead, numericalRecipes} to two decimal tolerance and employ a penalty function to enforce the fit range constraints. Additionally, the expansion, contraction, and shrink parameters are determined using a dimension dependent implementation \cite{Gao2010}. For the source term parameters $\alpha_e$ and $E_{cut}$, we see a general agreement with expectations as described above, albeit with a cutoff energy somewhat lower than the TeV level. The starlight component is also suppressedto the lower end of the allowed range, with a normalization factor $N_{SL} = 1.02 \times 10^{-12}$ which is a factor of $\sim 5$ lower than the value derived in section \ref{sec:ISRF}. We find a magnetic field value of $B_{\mu} = 1.7 \mu$G which is relatively small in comparison to those discussed in section \ref{sec:bfield}. While the actual structure of the field would involve higher central values with some spatially dependent fall-off, the average strength of the field over the space can be expected to take a smaller value. However, even with this in mind the magnetic field value is particularly low and likely not representative of the field strength within the inner regions of M31, especially the inner $\sim 1$ kpc where the synchrotron emission is calculated. Thus, we instead seek a configuration that allows for a higher magnetic field value.

\begin{table*}[tbph]
\centering
\def\arraystretch{1.5}
\begin{tabular}{cccccc}
	\hline\hline
	$\alpha_{e}$&\:\:$E_{cut}$ (GeV)&\:\: $N_{CR_e}$ (GeV$^{-1}$cm$^{-3}$s$^{-1}$) &\:\:$B_{\mu}$ ($\mu$G)&\:\: $N_{SL}$ &$ \:\: \chi^2_{min}/d.o.f$\\ 
	\hline
	$2.14$&$514$& $1.1 \times 10^{-25}$ &$1.7$& $1.02\times 10^{-12}$ & 3.32/7\\
	\hline\hline
	&&\multicolumn{2}{c}{Observational Values}&& \\
	\hline\hline
	$2.0-2.3$&$10^3-10^4$& --- &$ 5-10$& $ 10^{-12} - 10^{-11}$ & ---\\
	\hline\hline
\end{tabular}
\caption{\label{tab:PrimFitParams}Free parameters and their values in our best-fit model for a power law with exponential cutoff primary electron source. For reference, we have included in the bottom row the experimental values for the parameters as described in the text. Reference values for the source normalization are discussed in the context of CR power output in section \ref{sec:CRePower}.}
\end{table*}

\begin{figure}[h!]
\centering
\includegraphics[width=0.45\textwidth]{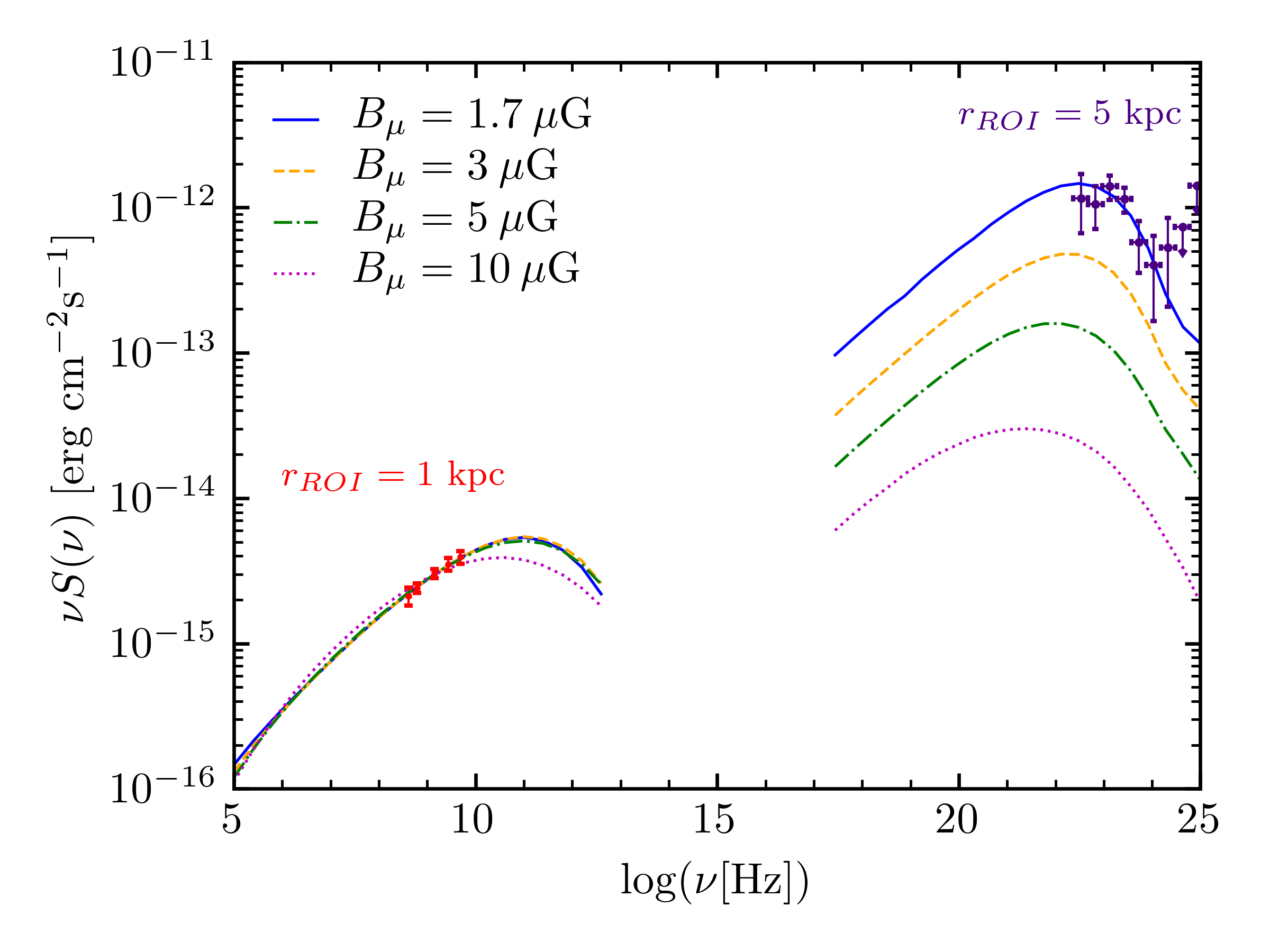}
\hfill
\caption{Spectrum due to synchrotron and IC emission from primary CRe for various values of the magnetic field, normalized to the observed radio emission. The best fit model is shown in green. Radio data are taken from \cite{WalterbosGrave} and gamma-ray data are taken from \cite{FermiM31}}
\label{fig:PrimB}
\end{figure}

One way in which we can potentially achieve a higher magnetic field  is to take into consideration the case where the radio emission is due to synchrotron radiation from cosmic-ray electrons, but the IC emission is not sufficient to recreate the Fermi observations, and remain agnostic as to the source of the gamma-ray emission. To do this, we increase the strength of the magnetic field and change $N_{CR_e}$ to reproduce the radio emission. In figure \ref{fig:PrimB} we show these fluxes for a few values of the magnetic field, and list the normalization factors in table \ref{tab:RadioNorm}. In this approach, we are essentially assuming that for reasonable magnetic field values the radio synchrotron emission in M31 can be produced predominantly by primary cosmic-ray electrons, while the source of the gamma ray emission remains unaccounted for. In later sections we use this approach in conjunction with comic-ray secondaries to account for the full spectrum of emission.

\begin{table}[tbph]
\centering
\def\arraystretch{1.5}
\begin{tabular}{cccccccc}
	\hline\hline
	$B_{\mu}$ ($\mu$G)&\:\: $N_{CR_e}$ (GeV$^{-1}$cm$^{-3}$s$^{-1}$) \\ 
	\hline
	3& $3.39 \times 10^{-26}$\\
	5& $1.28 \times 10^{-26}$\\
	10&$ 3.85 \times 10^{-27}$\\
	 \hline\hline
\end{tabular}
\caption{\label{tab:RadioNorm}Normalization factors for various magnetic field strengths in the case of CRe primaries, normalized to the radio emission.}
\end{table}

\subsubsection{Cosmic-ray Electron Power}\label{sec:CRePower}
\begin{figure}[th!]
		\centering
		\includegraphics[width=0.45\textwidth]{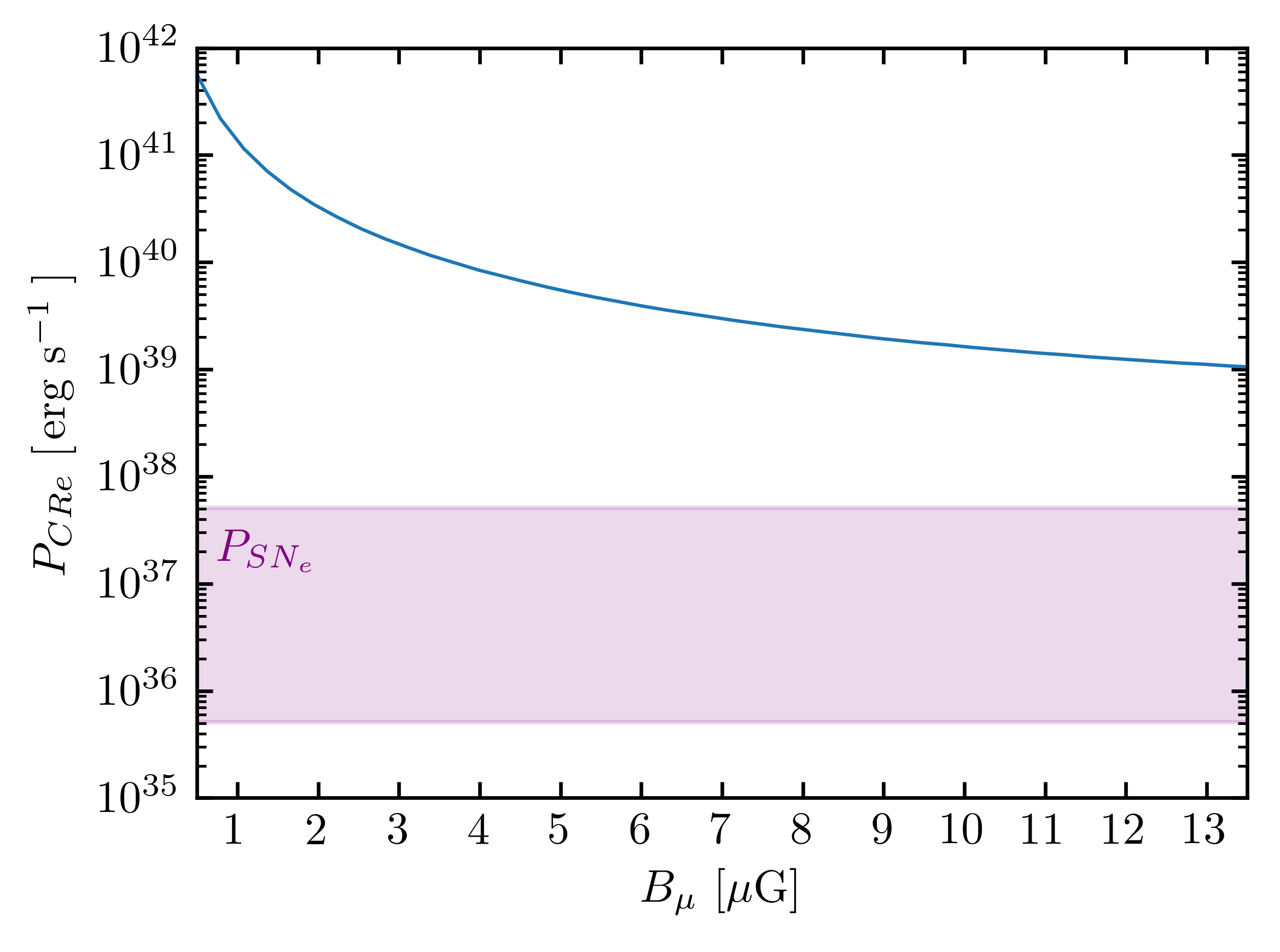}
		\hfill
		\caption{Power injection into CR$_e$ according to equation (\ref{eq:P_CRe}) for increasing magnetic field strength, normalized to the radio data. The purple region shows the range estimate for the electron power injection due to SNe as calculated using Eq. (\ref{eq:P_SNe}).}
		\label{fig:PvB}
\end{figure}

In order to place our value for the source term normalization $N_{CR_e}$ into a physical context, we can compare the total power output injected into cosmic-ray electrons with physical processes such as supernovae explosions. The power injected into the CRe for the source term of equation (\ref{eq:PrimSource}) is given by, 
\begin{equation}\label{eq:P_CRe}
P_{CR_e} = \int dV \int_{m_e}^{\infty}dE\: E \:Q_e(E),
\end{equation}
where $V$ is the diffusion volume. Meanwhile, the power injected into cosmic-ray electrons from supernovae is given by the expression, 
\begin{equation}\label{eq:P_SNe}
P_{SN, \:e} = \eta_e \Gamma E_{SN}
\end{equation}
where $\Gamma$ is the supernova (SN) rate, $E_{SN}$ is the total energy released in the SN explosion and $\eta_e$ is the efficiency of the SN energy transferred to the CRe. The SNe rate can be estimated from the observed star-formation rate (SFR), which in the case of M31 has been measured to be $\sim 0.2-0.4$ M$_{\odot}$ yr$^{-1}$ \cite{Ford2013, Rahmani2016}. Adopting a value of $SFR=0.25$ M$_{\odot}$ yr$^{-1}$ \cite{Ford2013}, the rate is then given by \cite{Horiuchi2011, Botticella2012}
\begin{equation}
\Gamma = SFR \times \frac{\int_{M_{min}}^{M_{max} }\psi(M) dM}{\int_{0.1M_{\odot}}^{100M_{\odot} }\psi(M)M dM}.
\end{equation}
We use the Salpeter initial mass function (IMF) \cite{Salpeter} defined over the main-sequence mass range of $0.1 -100\: M_{\odot}$ wherein $\psi (M) \propto M^{-2.35}$ and $\psi (M)dM$ gives the number of stars in the mass range $M+dM$. We take $M_{min}=8M_{\odot}$ and $M_{max}=40M_{\odot}$ in line with canonical CC SNe parameters \cite{Horiuchi2011}. This yields a SNe rate of $0.17$ per century, and the total energy output of for one supernova explosion is $E_{SN}\sim 10^{51}$ erg. While the efficiency at which energy is imparted to electrons during SN explosions is not well constrained, several estimates suggest values of $\eta_e =10^{-5} - 10^{-3}$ \cite{Tatischeff2009, Delahaye2010}. Putting these together, we obtain a lower limit on the power injected into CRe in SNe explosions to be $P_{SN,\:e} \approx 5.1\times 10^{35}$ erg s$^{-1}$, and an upper limit of $P_{SN, \:e} \approx 5.1\times 10^{37}$ erg s$^{-1}$. In figure \ref{fig:PvB} we show the power injected into CRe implied by our best-fit model while increasing the magnetic field and normalizing to the radio data (as in figure \ref{fig:PrimB}). We compare this with the estimated range of SNe power output for CRe and see that the necessary normalization to fit the radio data produces a power requirement that is substantially greater than the estimated SNe power budget for the lower magnetic fields, including at the best-fit value when also fitting the gamma-ray data at $B_{\mu} = 1.7\mu$G. Although the SNe power calculations involve a great deal of uncertainty, it is unlikely that the uncertainty is so great that it can be reconciled with the power output implied by our parameter model. Potential other cosmic-ray acceleration mechanisms such as PWNe could also contribute to the power total, however we can briefly demonstrate that this contribution is not enough to overcome the difference. For the case of pulsars, the relevant quantity is the spin down luminosity which can be expressed as
\begin{equation}
P_{PWN} = \frac{\eta W_0}{t_0 \left[1+ \left(\frac{t_{p}}{t_0}\right)\right]^2}
\end{equation}
where $\eta$ is the injection efficiency, $W_0$ is the pulsar energy output, $t_0$ is the typical pulsar decay timescale and $t_p$ is the pulsar lifetime \cite{Delahaye2010, malyshev, dellaTorre,lindenProfumo}. If we take as fiducial values, $\eta = 0.1$, $W_0 = 10^{50}$ erg, $t_{0} = 1$ kyr and assume $t_p \approx t_0$ we obtain a power contribution from pulsars of
\begin{equation}
P_{PWN} = \frac{\eta W_0}{4t_0 } \approx 8\times 10^{37} \text{erg s}^{-1}.
\end{equation}
While this value suggests that PWNe can contribute a significant amount of the CRe power, the estimate here is not sufficient to account for the necessary CRe power of our best-fit models, and thus does not have a significant impact on the results shown in figure \ref{fig:PvB}.
 
\subsection{\label{sec:resultsCRsecondary}Emissions from cosmic rays of hadronic origin}
We next consider the scenario in which the dominant contribution to the cosmic-ray population is in the form of primary cosmic-ray protons. Inelastic interactions between the CRp and the interstellar medium produce neutral and charged pions. The neutral pions decay into gamma-rays, while the charged pions decay into muons and neutrinos, which in turn decay into neutrinos and secondary CRe. This can be summarized as:
\begin{equation}
\begin{split}
&\pi^0 \rightarrow 2\gamma \\
&\pi^{\pm} \rightarrow \mu^{\pm} + \nu_{\mu}/\bar{\nu}_{\mu} \rightarrow e^{\pm} + \nu_{e}/\bar{\nu}_{e}
\end{split}
\end{equation}
For the most common astrophysical model of the CRp distribution we assume a simple power law:
\begin{equation}\label{eq:nCRp}
n_{CR_{p}}(E) = N_{CR_p} \left(\frac{E}{\text{GeV} }\right)^{-\alpha_{p}} 
\end{equation}
with $N_{CR_{p}}$ in units of GeV$^{-1}$ cm$^{-3}$. The resulting source terms for the gamma rays and cosmic-ray electrons have been previously calculated for this choice of CRp source distribution \cite{PfrommerEnsslin2004,MannheimSchlick1994,Schlickeiser2002}. Following \cite{PfrommerEnsslin2004} for the gamma-ray source term from $\pi^0$ decay yields the expression:

\begin{equation}\label{eq:QgammaPL}
\begin{split}
Q_{\gamma}(E,r) &=  N_{CR_p} \:n_{N}(r) c\sigma_{pp} \frac{4\xi^{2-\alpha_{\gamma}}}{3\alpha_{\gamma}}  \left(\frac{m_{\pi^0}}{\text{GeV}}\right)^{ -\alpha_{\gamma}}\\
&\times\left[ \left(\frac{2E_{\gamma}}{m_{\pi^0}}\right)^{\delta} +\left(\frac{2E_{\gamma}}{m_{\pi^0}}\right)^{ -\delta} \right]^{ -\alpha_{\gamma}/\delta}
\end{split}
\end{equation}
with, $\alpha_{\gamma } = 4/3\left(\alpha_{p} - 0.5\right)$, and the source for $e^{\pm}$ from the charged pion decay is given by 

\begin{equation}\label{eq:QePL}
Q_{e^{\pm}}(E,r) \simeq 2^6  N_{CR_p} \:n_{N}(r) c\sigma_{pp} \left(\frac{24 E}{\text{GeV}}\right)^{-\alpha_{\gamma}},
\end{equation}
as described in \cite{DolagEnsslin2000}. Here, $n_{N}(r)$ is the nucleon number density, which we take to be proportional to the thermal electron number density with $n_{N}(r) = \frac{1}{1-\frac{1}{2}X_{H} }  n_e(r)$ where $X_{H} = 0.24$ is the the primordial $\prescript{4}{}{\text{He}}$ mass fraction. The thermal electron density $n_e(r)$ can be modeled as a beta-fit of the form 
\begin{equation}
n_e  = n_{e,0} \left[1+\left(\frac{r}{r_c}\right)^2\right]^{-\frac{3}{2} \beta}
\end{equation}
with $\beta =0.49$ and $r_c = 54''$ \cite{LiuWang2010} and assuming $n_{e,0} \sim 0.1$ cm$^{-3}$ \cite{CaprioliElectron,CaprioliAmatoBlasiElectron}. The neutral pion mass is $m_{\pi^0} = 135$ GeV, $\xi$ gives the pion multiplicity taken to be $\xi =2$ for $\pi^0$, and $\sigma_{pp} =32$ mbarn is the proton collision cross-section. The shape parameter $\delta$ is given by $\delta =0.14\alpha^{-1.6}_{\gamma} + 0.44$. For this case, when fitting to both the radio and gamma-ray data using the same free parameters as the previous section (but with $E_{cut}$ excluded and $N_{CR_e}, \alpha_e$ replaced with $N_{CR_p}, \alpha_p$) we are unable to find a reasonable fit, due to the significant difference between the index required to fit the synchrotron emission to the radio simultaneously as the $\pi^0$ decay to the gamma-ray emission. Additionally, if we ignore the contribution from $\pi^0$ gamma rays and assume that synchrotron and IC emission from secondary electrons are dominantly responsible for the observed radio and gamma-ray emission we similarly do not find a good fit to the data.

\begin{figure}[h]
\centering
\includegraphics[width=0.45\textwidth]{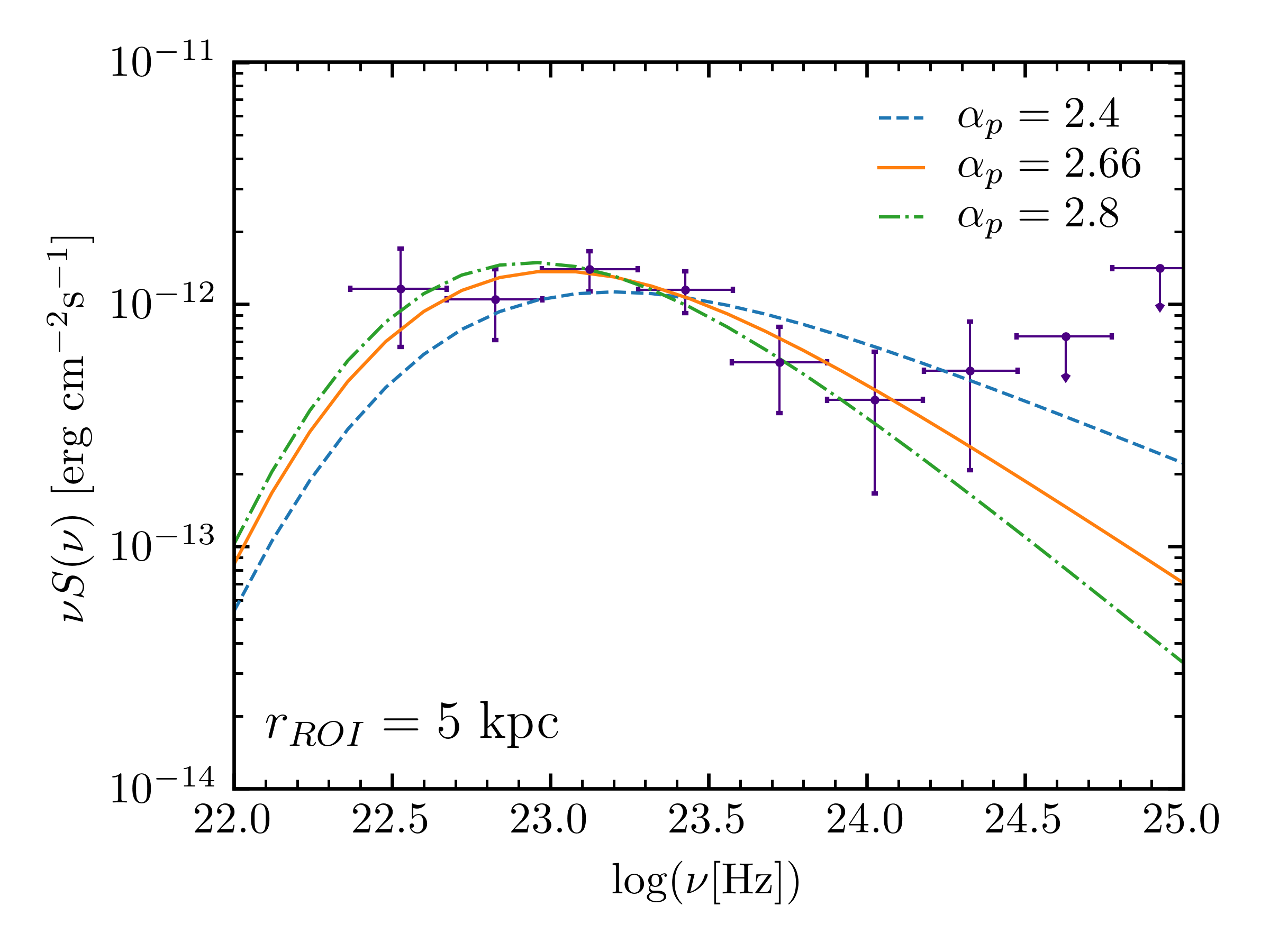}
\hfill
\caption{Spectrum due to $\pi^0$ decay for a few values of $\alpha_p$, normalized to the Fermi data from \cite{FermiM31}. }
\label{fig:pion_mult_alpha}
\end{figure}

\begin{figure}[h]
\centering
\includegraphics[width=0.45\textwidth]{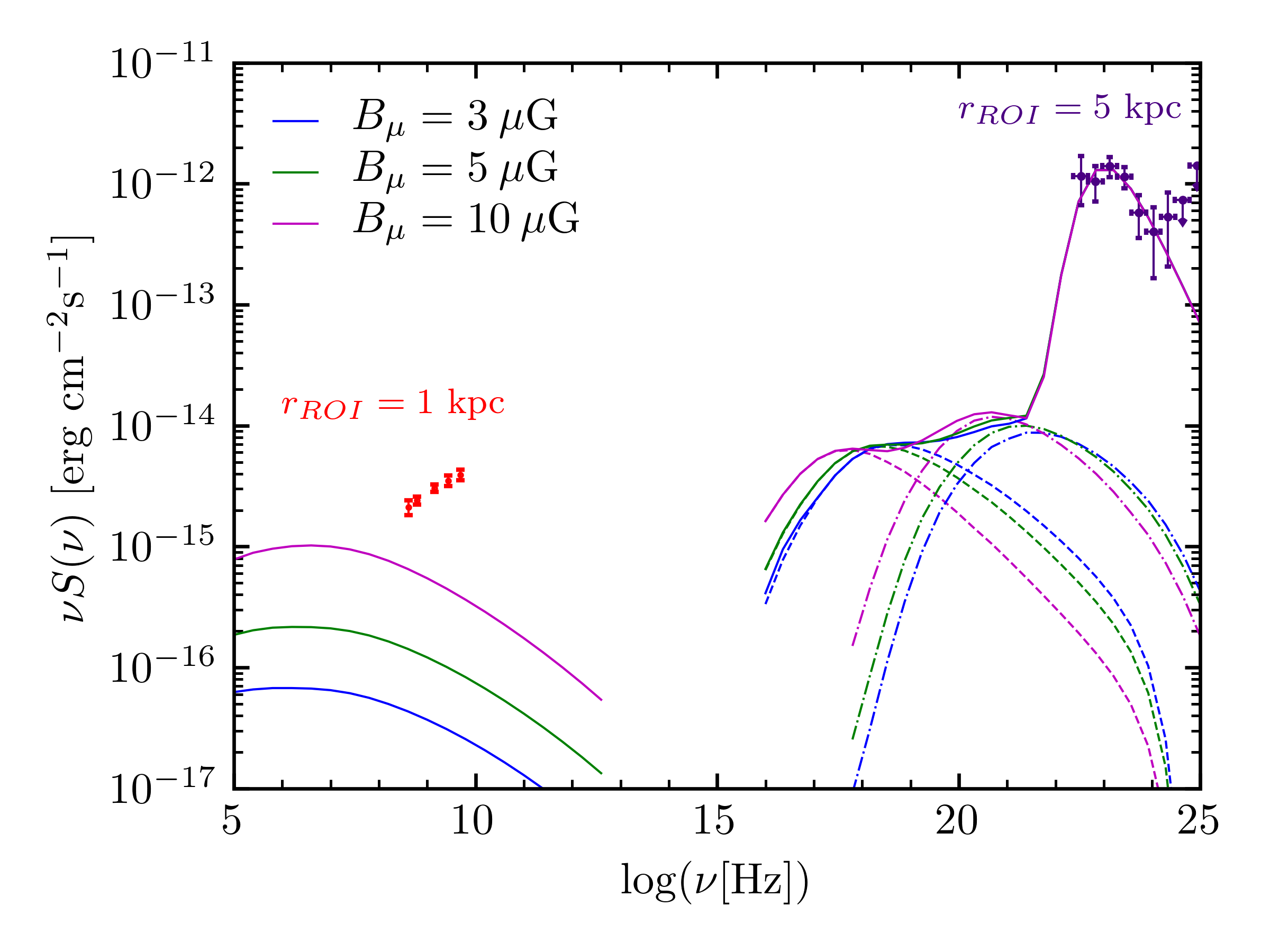}
\hfill
\caption{Emission due to decay of $\pi^{\pm},\pi^0$ into $e^{\pm}, \gamma$. Parameters were determined by fitting the pion decay gamma rays to the Fermi data with only the normalization $N_{CR_p}$ and the injection index $\alpha_p$ as free parameters, and for a selection of magnetic field strengths. The dashed lines are the CMB IC contribution, the dash-dot are the SL IC contribution, and the total emissions are the solid lines. Radio data are taken from \cite{WalterbosGrave} and gamma-ray data are taken from \cite{FermiM31}.}
\label{fig:SecBF}
\end{figure}

Instead, we determine $\alpha_p$ and $N_{CR_p}$ by only fitting the $\pi^0$ gamma rays to the Fermi data, while leaving the other parameters to be determined separately. With the only contribution to the fit being from the $\pi^0$ gamma ray contribution, we are find a best fit with $\alpha_p=2.66$ and $N_{CR_p} =8.89 \times 10^{-8}$ GeV$^{-1}$ cm$^{-3}$, also listed in table \ref{tab:SecFitPLParams}. The gamma-ray spectrum is shown in figure \ref{fig:pion_mult_alpha} along with a few other values of $\alpha_p$, normalized appropriately. In the selection of the remaining parameters that need to be determined (i.e. $B_{\mu}$ and $N_{SL}$) we are mainly constrained by the requirement that we are consistent with the field values in section \ref{sec:bfield} while not overproducing the radio emission, and $N_{SL}$ does not result in IC emission that significantly impacts the spectrum from pion decay gamma rays in the Fermi data energy range. In figure \ref{fig:SecBF} we show the result of this procedure with various values for the magnetic fields, and for simplicity a single starlight normalization $N_{SL} =5\times 10^{-12}$ in accordance with the discussion of section \ref{sec:ISRF}. In this figure, two things are evident: first, the gamma-ray emission provides a good fit to the Fermi data; and second, the spectral index required for this fit results in a significant mismatch to the radio data regardless of normalization or field strength. The index of the CRp distribution obtained is in agreement with other studies that suggest $\alpha_p \sim 2.5-2.75$ \cite{BlasiAmato2012,Huang2003, paglione1996}. 

\begin{table}[tbph]
\centering
\def\arraystretch{1.5}
\begin{tabular}{ccc}
	\hline\hline
	$\alpha_{p}$&\:\: $N_{CR_p}$ (GeV$^{-1}$cm$^{-3}$) & \:$\chi_{min}^2 / d.o.f$\\ 
	\hline
	$2.66$& $8.89\times 10^{-8}$& 2.25 / 5\\ 
	 \hline\hline
\end{tabular}
\caption{\label{tab:SecFitPLParams}Parameters and their values in our best-fit model for a power law primary proton source.}
\end{table}

\subsubsection{Cosmic-ray Proton Power}\label{sec:CRpPower}
As we did in section \ref{sec:CRePower}, we can again compare the power injected into CRp as implied by our fit parameters to the energy budget of SNe produced CRp. The power injection from SNe to CRp is of the same form as the CRe;
\begin{equation}\label{eq:P_SNp}
P_{SN,\:p} = \eta_p E_{SN} \Gamma_{SN}
\end{equation}
where the only difference is in the value of the power injection efficiency, $\eta_p$. While \cite{Tatischeff2009} inferred a value of $\eta_p \sim 10^{-5}-10^{-4}$, others have adopted higher values of $\eta_p \sim 10^{-3}$ \cite{CassemChenai}. Additionally, gamma ray observations suggest that up to $3-30\%$ of the SN kinetic energy can be imparted into the cosmic-ray protons \cite{FermiSN, thompson}. We therefore have quite a large range of possible values, finding  $P_{SN,\:p} \approx 5.1\times 10^{35}$ erg s$^{-1}$ for our lower bound and $P_{SN,\:p} \approx 1.53\times 10^{40}$ erg s$^{-1}$ as an upper bound.

\begin{figure}[th!]
		\centering
		\includegraphics[width=0.45\textwidth]{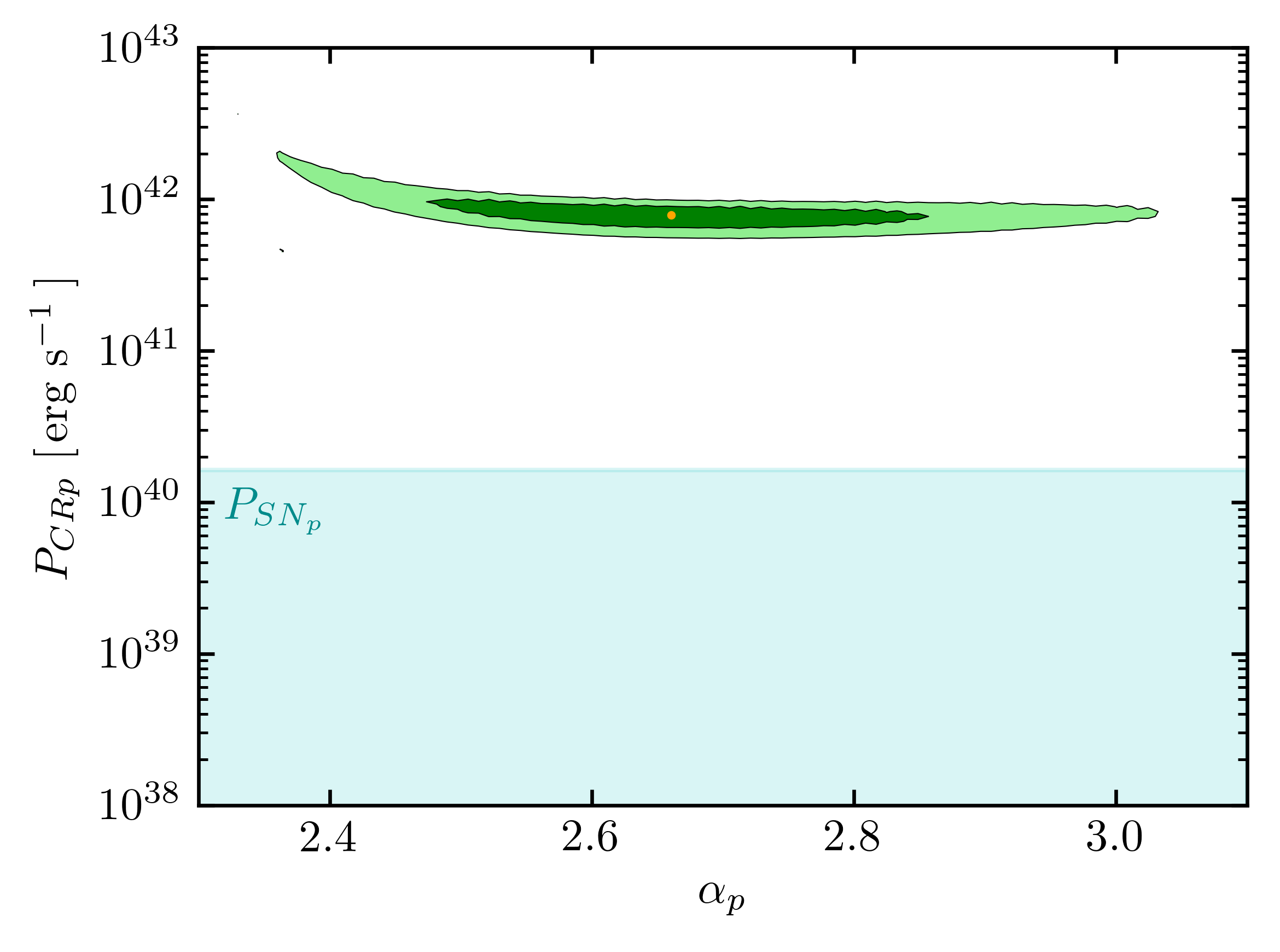}
		\hfill
		\caption{$1\sigma$ and $2\sigma$ contours of power injection into CR$_p$ according to equation (\ref{eq:P_CRp}) with $N_{CR_{p}}$ and $\alpha_{p}$ as free parameters and fitting only contributions from $\pi^0$ gamma rays. The light blue region shows a portion of the range estimate for the proton power injection due to SNe as calculated using equation (\ref{eq:P_SNp}). The best-fit point is given by the orange dot.}
		\label{fig:SecPCRp}
\end{figure}

To calculate the implied CRp power from our models, we take into account the diffusive properties of the CRp source distribution $n_{CR_p}$. Noting that for the heavier cosmic-ray protons the radiative energy losses of equation (\ref{eq:bloss}) are unimportant, we can consider only the propagation of the CRp by diffusion. The steady-state distribution of cosmic-ray protons has a characteristic diffusion timescale of $t_D(E) \approx r_h^2 /D(E)$ \cite{taylorIceCube,pavlidou, gaggero, longair, BlasiAmato2019}, which gives us an injection source term $n_{CR_p}(E) /t_D(E)$. We then have for the power injected into CRp:
\begin{equation}\label{eq:P_CRp}
P_{CR_p} = \int dV \int_{m_p}^{\infty}dE \:\left(E \frac{n_{CR_p}(E)}{t_D(E)} \right).
\end{equation}
In figure \ref{fig:SecPCRp} we show the contours of the implied power injected into CRp. In this case the power determined by the fit parameters still exceeds the estimated SNe power injection. This discrepancy between the SNe estimates and our calculated power is not as extreme as in the primary $P_{CR_e}$ scenario for lower magnetic fields, though for higher field values, the CRe power is just over an order of magnitude greater than the upper SNe power, as opposed to the almost two order of magnitude difference for the CRp seen here.

\subsection{\label{sec:resultsMultiCRsources}Multi-component cosmic ray source model}
\begin{figure}[h!]
\centering
\includegraphics[width=0.5\textwidth]{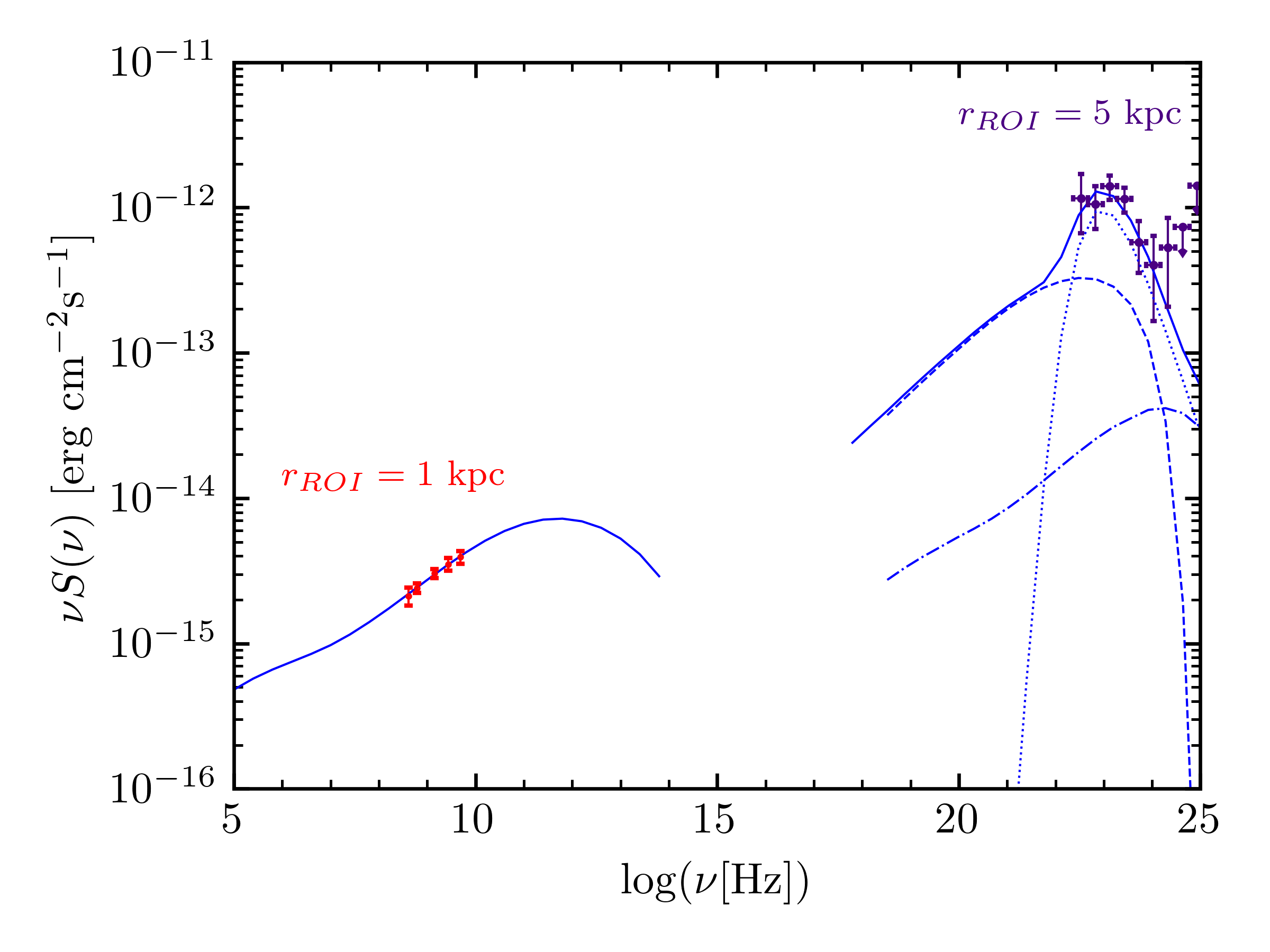}
\hfill
\caption{ SED of the best-fitting multi-component CR model, including contributions from primary CRe, as well as secondary CRe and gamma rays of hadronic origin. Fit parameters are listed in the top row of} Table \ref{tab:MCFitParams}. The dashed lines are the IC CMB contribution, the dash-dot are the IC SL contribution, dotted lines are the $\pi^0$ gamma-rays, and the total emissions are the solid lines. Radio data are taken from \cite{WalterbosGrave} and gamma-ray data are taken from \cite{FermiM31}
\label{fig:MCmodel}
\end{figure}

In the previous sections we were working under the assumption that the cosmic-ray source was dominated by either primary production of CRe or hadronic production of secondary CRe and $\pi^0$ gamma rays. However, another possible scenario would be where both of these cosmic-ray production mechanisms are incorporated. We can therefore consider a multi-component model that includes contributions of both the primary source as well as the hadronically produced sources. For the gamma-ray source term, the only contribution is from the decay of pions produced in inelastic hadronic collisions as described by (\ref{eq:QgammaPL}). The electron source term for the multi-component model is the sum of the source terms in equations (\ref{eq:PrimSource}) and (\ref{eq:QePL}):
\begin{equation}
\begin{split}
Q^{MC}_{e^{\pm}}(E,r) &= N_{CR_e} \left(\frac{E}{\text{GeV} }\right)^{-\alpha_{e}} e^{-E/E_{cut}} \\
&+ 2^6  N_{CR_p} \:n_{N}(r) c\sigma_{pp} \left(\frac{24 E}{\text{GeV}}\right)^{-\alpha_{\gamma}} 
\end{split}
\end{equation}
with, $\alpha_{\gamma } = 4/3\left(\alpha_{p} - 0.5\right)$. The best-fit results are listed in table \ref{tab:MCFitParams} along with a selection of parameter sets with fixed magnetic fields or fixed $N_{SL}$. The SED for the best fit is shown in figure \ref{fig:MCmodel}. 

\begin{table*}[tbph]
\centering
\def\arraystretch{1.5}
\begin{tabular}{cccccccccc}
	\hline\hline
	$\alpha_{e}$&\:\:$\alpha_{p}$&\:\:$E_{cut}$ (GeV)&\:\: $N_{CRe}$ (GeV$^{-1}$s$^{-1}$cm$^{-3}$) &\:\: $N_{CRp}$ (GeV$^{-1}$cm$^{-3}$) & \:\:$B_{\mu}$ ($\mu$G)&\:\: $N_{SL}$ &\:\: $\chi^2_{min}/d.o.f$\\ 
	\hline
	$2.04$&\:\:$2.75$&$1658$& $1.32\times 10^{-26}$&$7.48\times 10^{-8}$& $4.8$& $1.10 \times 10^{-12}$ & $2.27/5$\\ 
	$2.09$&\:\:$2.75$&$854$& $3.29\times 10^{-26}$&$5.24\times 10^{-8}$& $\prescript{a}{}3$& $1.02 \times 10^{-12}$ & $2.41/6$\\ 
	$1.92$&\:\:$2.71$&$1633$& $6.17\times 10^{-27}$&$7.75\times 10^{-8}$& $\prescript{a}{}7$& $1.12 \times 10^{-12}$ & $2.42/6$\\ 
	$1.71$&\:\:$2.67$&$1550$& $2.73\times 10^{-27}$&$7.01\times 10^{-8}$& $\prescript{a}{}10$& $1.44 \times 10^{-12}$ & $2.80/6$\\ 
	$1.57$&\:\:$2.70$&$1353$& $1.77\times 10^{-27}$&$7.13\times 10^{-8}$& $12.4$& $\prescript{a}{}5 \times 10^{-12}$ & $3.25/6$\\ 
	 \hline\hline
\end{tabular}
\caption{\label{tab:MCFitParams}Parameters and their values in a selection of well fitting models for the multi-component model, along with the corresponding $\chi^2_{min}$. The best-fit model parameters are given in the top row and the corresponding SED is plotted in figure \ref{fig:MCmodel}.} Parameters that are held fixed in a given model are denoted with the $a$ prescript.
\end{table*}

\begin{figure}[h!]
\centering
\includegraphics[width=0.5\textwidth]{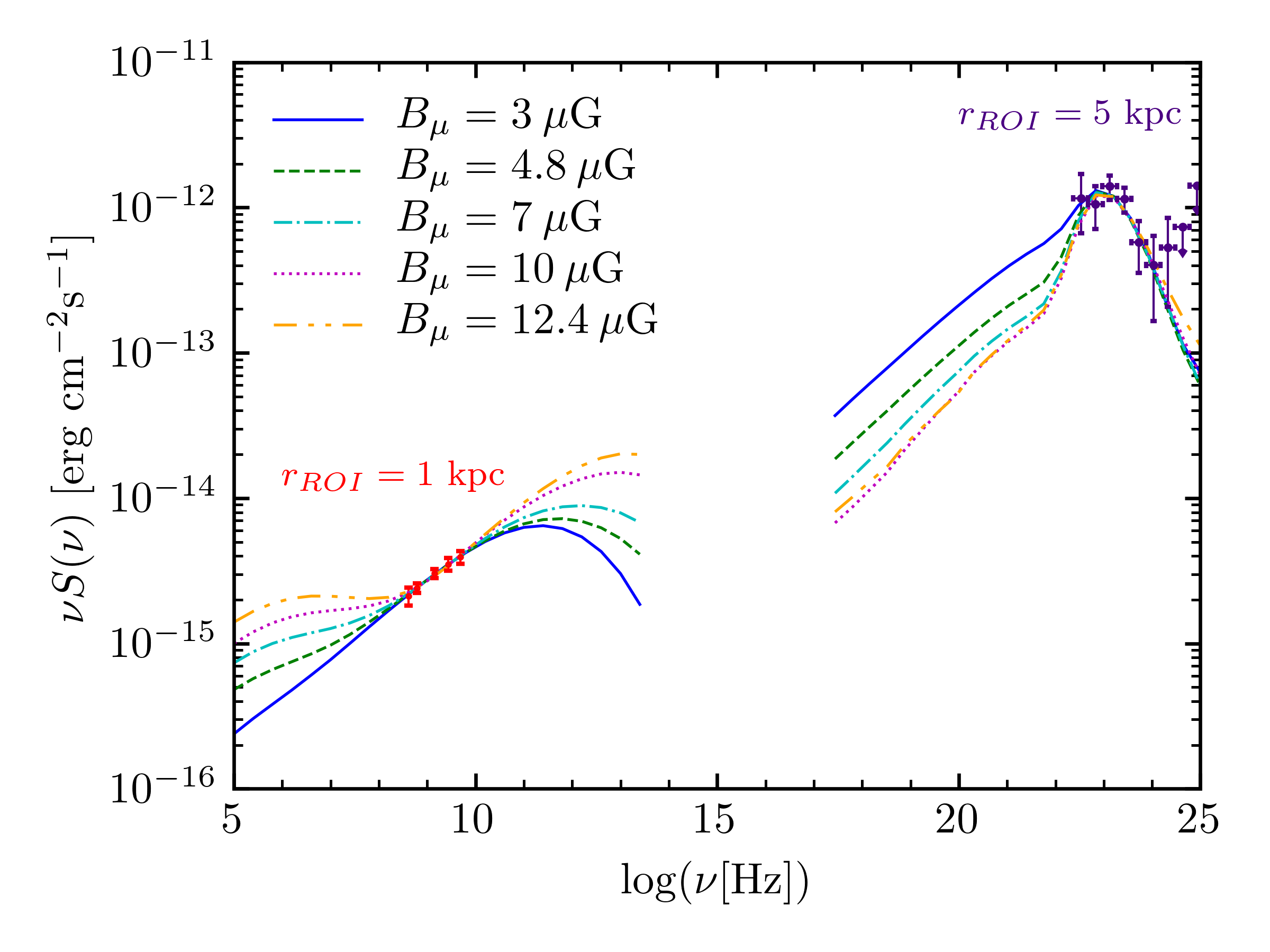}
\hfill
\caption{SED of the multi-component CR model for each of the parameter sets in table \ref{tab:MCFitParams}. Radio data are taken from \cite{WalterbosGrave} and gamma-ray data are taken from \cite{FermiM31}}
\label{fig:MC_Mult}
\end{figure}

For the injection indices, we obtain values of $\alpha_e = 2.04$ and $\alpha_p = 2.75$, which are both within the ranges discussed in sections \ref{sec:resultsCRPrimary} and \ref{sec:resultsCRsecondary}, while the normalization factors $N_{CRe}$ and $N_{CRp}$ do not deviate significantly from the values found in those sections. The cutoff energy $E_{cut} = 1658$ GeV is in line with the $\sim$ TeV level used in previous cosmic-ray studies \cite{Delahaye2010, DiMauro2014, manconi2017, Fang2017,bernardo2013}, but higher than in the primary-only case. The magnetic field is also higher here than in the primary only case and is in good agreement with M31 magnetic field estimates. The similarity between the parameters of the multi-component model  and the primary-only or secondary-only models is reflected in that for the multi-component model each of the two components (primary and secondary) have separate regimes of dominance. That is to say, the radio is predominantly due to the primary CRe whereas the gamma-rays are mainly due to the neutral pion decay gamma rays. This resolves the discrepancy in the model with purely hadronically produced CRe between the spectrum of radio data and the predicted synchrotron emission. 
In addition to the best-fit model, we also list in table \ref{tab:MCFitParams} models in which we hold the magnetic field fixed and fit for the remaining free parameters. We do this as well where we instead hold $N_{SL}$ fixed to the value discussed in \ref{sec:ISRF} and fit the remaining parameters. In either case, the parameter held fixed is denoted in table \ref{tab:MCFitParams} by the `$a$' superscript. The spectra for each model in \ref{tab:MCFitParams} are plotted in figure \ref{fig:MC_Mult}. With different field strengths we are still able to find good fits to the data, with only very slight changes to the $\chi_{min}^2$. This suggests that in the multi-component model there is no issue with a suppressed magnetic field as in the primary-only case. Again we see that the starlight normalization is highly suppressed and the IC emission is heavily dominated by the CMB component. Since this appears to be a fairly extreme scenario for the central region of the galaxy, we try to achieve a more reasonable value by holding the normalization fixed at $N_{SL} = 5\times10^{-12}$ as derived in section \ref{sec:ISRF} and fit the remaining parameters. We are still able to achieve a good fit, however it requires a relatively higher magnetic field of $B_{\mu} = 12.4 \mu$G in order to suppress the stellar IC component, as well as a low injection index of $\alpha_e =1.57$. 

\begin{figure}[h!]
\centering
\includegraphics[width=0.5\textwidth]{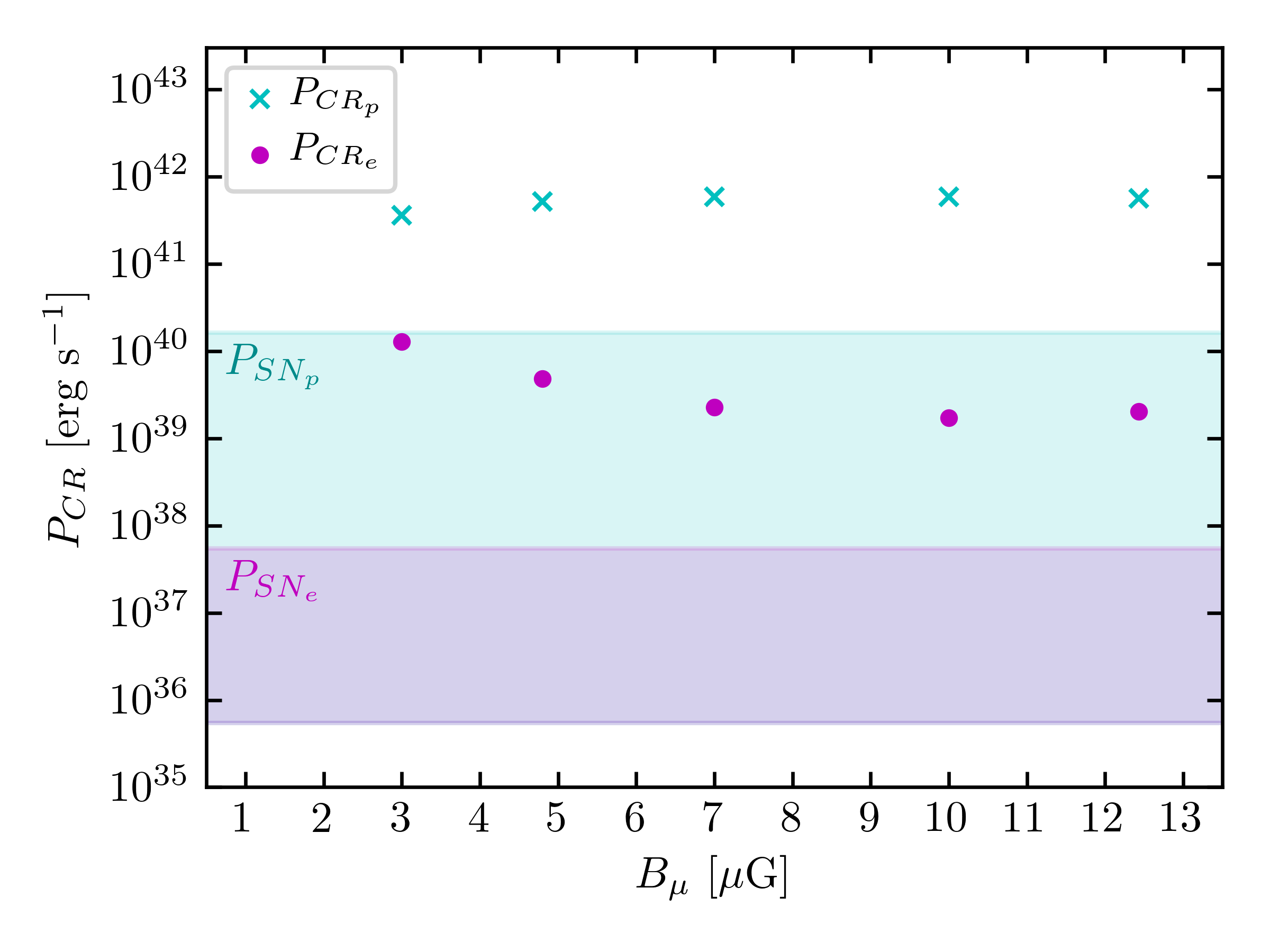}
\hfill
\caption{Power injection into CRe and CRp for each of the models in table \ref{tab:MCFitParams}, plotted against magnetic field. Note that the region of SNe power into CRp (cyan) fully overlaps the SNe power injection into CRe region (magenta).}
\label{fig:MC_Ppoints}
\end{figure}

We once again compare the power injection into CRe and CRp implied by the parameters of our fit with the estimated SN injected power. Noting that the source term parameters for the CRe and CRp do not deviate significantly from the values found in section \ref{sec:resultsCRPrimary} and \ref{sec:resultsCRsecondary}, similar results in this comparison can be expected here. In fact, that is essentially what we see in figure \ref{fig:MC_Ppoints}, wherein we show the implied CR power from our models for the various magnetic field values compared with the SN power injection estimates of sections \ref{sec:CRePower} and \ref{sec:CRpPower}. We see that the implied CRe power injection decreases for models with higher magnetic field (cf. figure \ref{fig:PvB}), while the CRp injection remains relatively constant with some slight increase due to suppression of the primary CRe induced IC emission. However, neither are within their respective ranges for the SN source power.  Although there is a discrepancy between the implied power injection of our cosmic-ray parameter sets and the estimated supernovae contribution, the great deal of uncertainty in the SNe power estimates makes it difficult to make concrete statements on the viability of these models on this basis alone.

\section{\label{sec:xraydiffuse}Diffuse X-ray Emission in M31}
While the focus of this analysis has been on the radio and gamma-ray emission, X-ray emission in M31 provides another potential avenue to study. However, several observations of the X-ray emission in the bulge of M31 have detected the presence of an unresolved diffuse component, using data from {\it ROSAT} \cite{Supper01, Supper1997}, {\it XMM-Newton} \cite{Shirey2001}, and {\it CHANDRA} \cite{LiWang2007}. In each of these studies, diffuse X-ray flux in the inner $\sim$ 1 kpc of M31 is observed at a flux level of falling roughly between $\sim 3 - 5 \times 10^{-12}$ erg cm$^{-2}$ s$^{-1}$, and can likely be attributed to the presence of thermal hot gas and unresolved X-ray point sources. We note that this observed X-ray emission within 1 kpc  has a higher flux than the X-rays produced in any of our cosmic-ray models from the previous sections, even despite the computed X-ray emission being within a 5 kpc radius. We thus conclude that for these cosmic ray models and our astrophysical setup the X-ray emission in M31 does not provide particularly useful information due to the bright diffuse emission in the bulge of M31 being considerably brighter than what we would obtain in our models. 

\section{\label{sec:conclusion}Conclusion}
We have examined the possibility of a cosmic-ray origin for the multi-wavelength emission in the Andromeda galaxy, specifically addressing the origin of the recently detected extended gamma-ray emission. We considered three models for the production of cosmic rays. First, we considered a primary injection of CRe obeying a power law with an exponential cutoff, then considered production of secondary CRe and gamma rays produced from interactions of a power law distribution of primary CRp. Finally, we looked at a multi-component model that incorporates both of these cosmic-ray sources. We then fit the synchrotron and IC fluxes arising due to the presence of the primary and secondary CRe, as well as the gamma ray emission from neutral pion decay, to available radio data and a recent Fermi gamma-ray detection in M31.

For the primary CRe scenario, we find best fit parameters for the injection spectrum $\alpha_e = 2.14$ and cutoff energy  $E_{cut} = 514$ GeV. The injection index is consistent with expected values for CRe sources such as SNR. The cutoff energy is slightly lower than expected, however not wholly inconsistent with expected values on the order of TeV. The magnetic field value of $B_{\mu} = 1.7\: \mu$G and the starlight normalization are both suppressed in the fit. We also considered higher magnetic fields and renormalized the synchrotron emission to match the radio data. This suppresses the IC gamma-ray emission, requiring that we account for the Fermi data separately which was done in the multi-component model. We then compared the power injection into CRe implied by our model with the expected range of power injection due to SNe. We saw that even by increasing the magnetic field in order to lower the normalization constant $N_{CRe}$, the power injection implied by our models was well above the expected output from astrophysical sources such as PWNe and SNe.

In the case where we considered contributions from  only secondary cosmic rays of hadronic origin, we were unable to find a good fit to both the radio and gamma-ray data simultaneously. Rather, we assumed that the gamma rays were purely from the neutral pion decay and found a CRp distribution index of $\alpha_p = 2.66$, consistent with previous results for $\pi^0$ gamma-ray studies, along with a CRp distribution coefficient of $N_{CR_p} = 8.89\times 10^{-8}$ GeV$^{-1}$ cm$^{-3}$. With this arrangement we then manually selected the magnetic field and starlight energy density, and found that for a variety of field strengths the calculated flux remains below the radio data, and even for a higher selected value of $N_{SL} = 5\times 10^{-12}$ there was no conflict between the IC emission and the gamma-ray data. We again compared the power injection into CRp from SNe with the implied power output of our models, and found that the CRp injection is also greater than the estimated SNe output.

Finally, we consider a combined ``multi-component model'' that incorporates the contributions from both the primary CRe as well as the secondary CRe of hadronic origin. Although here the power budget concerns remained due to minimal variation in the best fit normalization constants, this scenario gives the best overall fit to the data, while still providing similar parameter values as in the primary-only and secondary-only cases. We found the best-fit $\alpha_e=2.04$ and $\alpha_p=2.75$, both similar to the values discussed in sections \ref{sec:resultsCRPrimary} and \ref{sec:resultsCRsecondary} respectively, while the best-fit magnetic field was found to be $B_{\mu} =4.8 \: \mu$G and $E_{cut}$ was $1658$ GeV. Additionally, the multi-component model offers a large degree of flexibility in the parameter choices, as evidenced by good fits for a range of multiple magnetic fields values and $E_{cut}$ on the order of a few TeV, as well as for higher $N_{SL}$ values in accordance with the observed stellar luminosity in the central region of M31. In our final power comparison we saw similar results as in the primary-only and secondary-only scenarios. That is, both the implied CRe and CRp power in our models were greater than the estimated power output from astrophysical sources, and this held at a wide range of magnetic fields values. This suggests that although the spectra can be fit well with a multi-component model, the input power needed for the cosmic-ray sources is consistently more than an order of magnitude above what is expected from supernova as galactic cosmic-ray accelerators.
Furthermore, as mentioned in the introduction and discussed more thoroughly in the original detection paper \cite{FermiM31} along with earlier and subsequent Fermi M31 studies \cite{Pshirkov,EcknerMSP,FragioneMSP}, the gamma-ray emission does not appear to correlate with star-formation or gas-rich regions. CRp produced at larger radii that then diffuse into the emission region may contribute to the observed signal, although this does not address the lack of gas for the CRp to interact with in the interior regions of the galaxy. Another possibility is that the CRp are remnants from a previous period of higher star-formation. However, the stellar population of the bulge is dominated by stars with ages $\gtrsim 4-12$ Gyr \cite{Olsen2006,Saglia2010,Dong2018}, compared to a CRp escape time of $\sim 10-100$ Myr, which suggests that the majority of CRp would have likely left the system in the time since this higher star-formation activity in M31. This morphological point along with the power discrepancy combine to disfavour a purely CR explanation, particularly one that relies on $\pi^0$ gamma-rays from CRp to explain the observed gamma-ray emission.
\begin{acknowledgments}
This study is based on work supported by the National Science Foundation under Grant No. 1517545. A.M. is supported by a Department of Education GAANN fellowship. S.P. is partly supported by the US Department of Energy, grant number DE-SC0010107. 
\end{acknowledgments}
\bibliographystyle{unsrtnat}
\bibliography{ref_CR}
\end{document}